%% file: cas-dc-template.tex
\documentclass[a4paper,fleqn]{cas-dc}
\usepackage[caption=false,font=footnotesize]{subfig}
\usepackage{natbib}
\usepackage{multicol}
\usepackage{wrapfig}

\usepackage{amssymb, amsmath, mathtools,url}
\usepackage{graphicx,epsfig,psfrag,color}
\usepackage{siunitx}
\usepackage{booktabs}
\usepackage{algorithm2e}
\usepackage[long]{optidef}

\usepackage{comment}
\usepackage{hyperref}
\usepackage[capitalise]{cleveref}
\DeclareMathOperator*{\minimize}{minimize}
\usepackage{graphicx}
\newtheorem{prop}{Proposition}
\usepackage[nonumberlist]{glossaries}
\setglossarystyle{superheaderborder}
\newglossaryentry{xr}
{
    name=$x_{ij}^{\text{r}}(t)$ ,
    description={A decision variable for the number of vehicles to rebalance from station $i$ to station $j$ at time interval $t$}
}
\newglossaryentry{xc}
{
    name=\textbf{$x_{ij}^{\text{c}}(t)$} ,
    description={he number of vehicles to drive customers from station $i$ to station $j$ at time interval $t$}
}
\newglossaryentry{s}{
name ={\textbf{$s_{ij}(t)$} },
description={A decision variable for the imbalance, i.e. describes how many customers to not pick-up at time $t$ that wants to go from station $i$ to station $j$}
}
\newglossaryentry{lambda}{
name ={\textbf{$\lambda_{ij}(t)$} },
description={The number of customers that wants to travel from station $i$ to station $j$ at time $t$}
}
\newglossaryentry{deltat}{
name ={\textbf{$\Delta t$}},
description={ The discrete time interval length}
}
\newglossaryentry{cr}{
name ={\textbf{$c_{ij}^{\text{r}}(t)$} },
description={The cost of rebalancing one vehicle from station $i$ to station $j$ at time $t$}
}
\newglossaryentry{cl}{
name ={\textbf{$c_{\lambda}$}},
description={ The cost of leaving out one customer that wants to go from station $i$ to station $j$ at time $t$}
}
\newglossaryentry{kappa}{
name ={\textbf{$\kappa_{ij}(t)$}},
description={The travel time, in discrete time intervals, to drive from station $i$ to station $j$ at time $t$}
}
\newglossaryentry{phi}{
name ={\textbf{$\phi_{i}(t)$}},
description={Initial number of idle vehicles in each station. }
}
\newglossaryentry{theta}{
name ={\textbf{$\Theta \subset \mathbb{R}^2$} },
description={Operating area for vehicles}
}
\makenoidxglossaries
\begin{document}
\let\WriteBookmarks\relax
\def\floatpagepagefraction{1}
\def\textpagefraction{.001}
\shorttitle{A Predictive Chance Constraint Rebalancing Approach to Mobility-on-Demand Services}
\shortauthors{S.E.T. Jacobsen, B. Kulcs\'ar, A. Lindman}
\title[mode = title]{A Predictive Chance Constraint Rebalancing Approach to Mobility-on-Demand Services}                     
\tnotemark[1]


\author[1,2]{{Sten Elling} Tingstad Jacobsen}[]
\ead{stingsta@volvocars.com}

\cormark[1]
\cortext[cor1]{Corresponding author}

\author[2]{Anders Lindman}[style=chinese]
\author[1]{Bal\'azs Kulcs\'ar}[]
\ead{kulcsar@chalmers.se}
\address[1]{Department of Electrical Engineering, Chalmers University of Technology,  Gothenburg, Sweden}

\affiliation[2]{organization={Volvo Cars AB},
    city={Gothenburg},
    country={Sweden}}

\begin{abstract}
This paper considers the problem of supply-demand imbalances in Mobility-on-Demand (MoD) services, such as Uber or DiDi Rider. Such imbalances are due to uneven stochastic travel demand and can be prevented by proactively rebalance empty vehicles to areas where the demand is high. To this end we propose a method that include estimated stochastic travel demand patterns into stochastic model predictive control for rebalancing of empty vehicles in a autonomous MoD ride-hailing service, where the objective is to minimize the imbalance and the rebalance distance driven. More precisely, we first estimate passenger travel demand using Gaussian Process Regression (GPR), which provides demand uncertainty bounds for time pattern prediction. We then formulate a stochastic model predictive control for the autonomous ride-hailing service and integrate demand predictions with uncertainty bounds into a receding horizon MoD optimization. In order to guarantee constraint satisfaction in the above optimization under estimated stochastic demand prediction, we employ a probabilistic constraining method with user defined confidence interval. Receding horizon MoD optimization with probabilistic constraints thereby calls for Chance Constrained Model Predictive Control (CCMPC). The benefits of the proposed method are twofold. First, travel demand uncertainty prediction from data can naturally be embedded into the MoD optimization framework. We show that for a given minimal fleet size the imbalance in each station can be kept below a certain threshold with a user defined probability. Second, CCMPC can further be relaxed into a Mixed-Integer-Linear-Program (MILP) and we show that the MILP can be solved as a corresponding Linear-Program which always admits a integral solution. Finally, we demonstrate through high-fidelity transportation simulations, that by tuning the confidence bound on the chance constraint close to optimal oracle performance can be achieved. The corresponding median customer wait time is reduced by $4\%$ compared to using only the mean prediction of the GPR.
\end{abstract}

\maketitle
\input{sections/introduction.tex}
\input{sections/model.tex}
\input{sections/problemStatement.tex}
\input{sections/simulationResults.tex}

\input{sections/conclusion.tex}

\input{sections/acknowledgment.tex}

\appendix
\glsaddall
\printnoidxglossaries
\section{Total Unimodular}\label{app:TU}
There are certain cases where the optimal solution of the LP relaxation of an MILP is guaranteed to be integral. Consider the following MILP,
\begin{equation}
\begin{aligned}
&\minimize_{x} \quad & c^{\top}\mathbf{x} \\
& \text{subject to} \notag \\
& \quad & A \mathbf{x}\leq b \\
& \quad & \mathbf{x} \in \mathbb{N}.
\end{aligned}
\end{equation}
If the A matrix is totally unimodular (TU) then the LP relaxation will always have one integral solution \citep{Hoffman2010}. A matrix A is TU if the determinant of A is $\pm 1$. We will now prove that \cref{eq:chanceconstraintgpr} have this property. Let $\mathbf{x}$ be the vector of all decision variables $s_{ij}(t)$, $x^{\text{c}}_{ij}(t)$ and $x^{\text{c}}_{ij}(t)$. Since the decision variables in \cref{eq:ccgprim,eq:1imb2,eq:1flow} enters as a subtraction or addition all entries i A will either be 1, -1 or 0. 
\begin{prop}
    Let $\mathbf{A} \in \{-1,0,1\}^{n\times m}$. If every column of A has at most one 1 and at most one -1, then A is totally unimodular \citep{SEYMOUR1980305}.
\end{prop}
Since $x^{\text{c}}_{ij}(t)$ is the only decision variable that appears both in \cref{eq:ccgprim} and \cref{eq:1flow}, but with different sign, each column of A will consist of at most one 1 and one -1 and hence the A matrix is TU. Therefore the MILP \cref{eq:chanceconstraintgpr} can be solved as a linear program (LP).



\bibliographystyle{agsm}
\bibliography{ref_journal}

\vspace{1cm}
\section*{Authors biography}
\begin{wrapfigure}{l}{25mm}
    \includegraphics[width=1.2in,height=1.4in,clip,keepaspectratio]{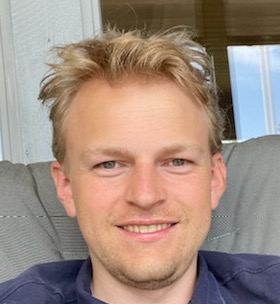}
\end{wrapfigure}
\noindent \textbf{Sten Elling Tingstad Jacobsen} received the M.S. degree in applied physics from Chalmers University of Technology, Sweden, in 2020. He is currently working on his Ph.D. jointly at Volvo Cars AB and the Department of Electrical Engineering, Chalmers University of Technology, Sweden. His research interests include optimization, machine learning, control, and their application to transportation systems and the automotive area. \par
\newpage
\vspace{1cm}

\begin{wrapfigure}{l}{25mm} \includegraphics[width=1.2in,height=1.4in,clip,keepaspectratio]{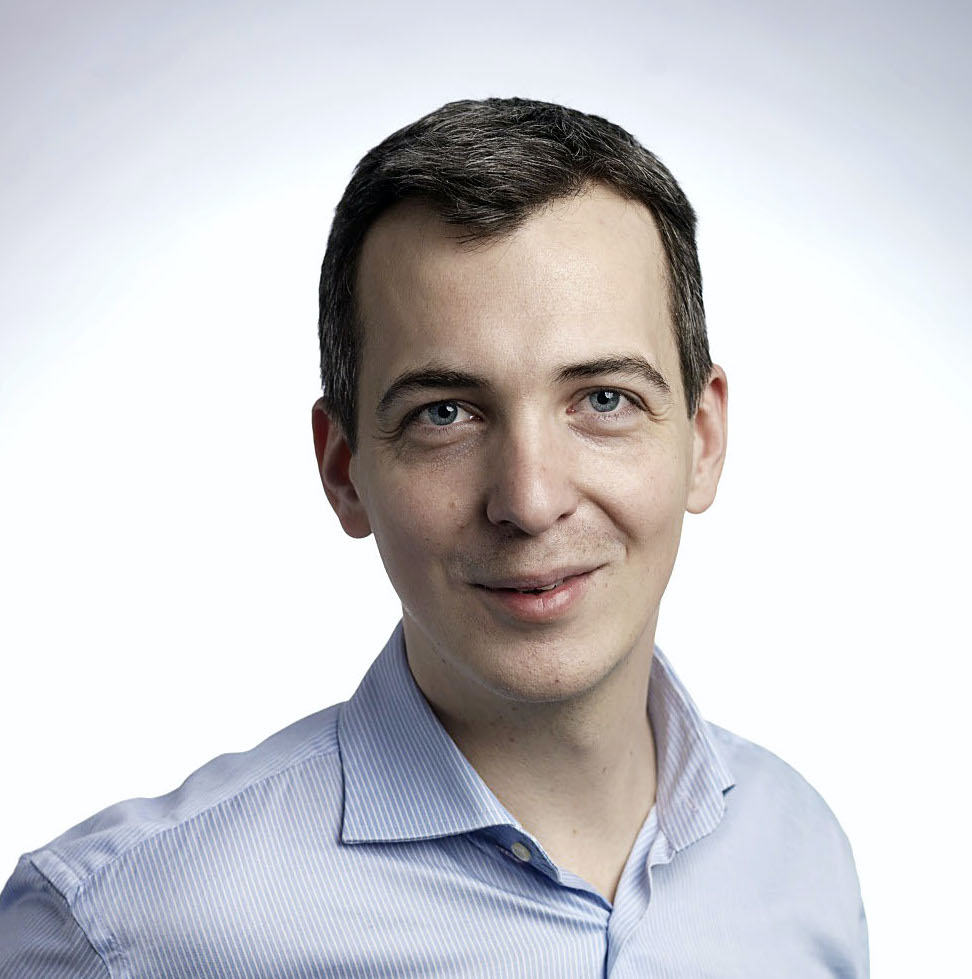}
\end{wrapfigure}
\noindent \textbf{Balázs Kulcsár} received the M.Sc. degree in traffic engineering and the Ph.D. degree from Budapest University of Technology and Economics (BUTE), Budapest, Hungary, in 1999 and 2006, respectively. He has been a Researcher/Post-Doctor with the Department of Control for Transportation and Vehicle Systems, BUTE, the Department of Aerospace Engineering and Mechanics, University of Minnesota, Minneapolis, MN, USA, and with the Delft Center for Systems and Control, Delft University of Technology, Delft, The Netherlands. He is currently a Professor with the Department of Electrical Engineering, Chalmers University of Technology, Göteborg, Sweden. His main research interest focuses on traffic network modeling and control.

\vspace{1cm}
\begin{wrapfigure}{l}{25mm} 
    \includegraphics[width=1.2in,height=1.4in,clip,keepaspectratio]{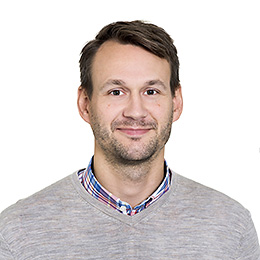}
\end{wrapfigure}
\noindent
\textbf{Anders Lindman} received his Ph.D. (2017) in Computational Materials Physics from Chalmers University of Technology (Sweden). He joined Volvo Car Corporation in 2018 where he is active in the areas of Vehicle Energy Efficiency and Future Mobility.

\end{document}

%% file: sections/introduction.tex

\section{Introduction}



The fast growing urbanization in the world puts major challenges on urban transportation \citep{JRC116644}. In Europe the urbanization is expected to grow from $74\%$ in 2018 to $84\%$ in 2050 \citep{united2019world}. Traditionally, urban transportation is improved by infrastructure investments in road expansions and public transportation. However, with recent development of new technologies within automation, connectivity, electrification and shared services, there is a potential for new transport solutions to satisfy the increasing demand. Transportation modes that have gained huge interest and market share are mobility-on-demand (MoD) services, e.g. mobility-as-a-service providers and car-rental pools  \citep{Zardini_2021}. These types of services are more flexible than public transportation and can in fact complement it \citep{amod_publictrans}. Moreover, the combination of MoD and autonomous vehicles (AVs), Autonomous Mobility-on-Demand (AMoD), has been at the center of research for over a decade \citep{Zardini_2021}. AMoD is predicted to become one of the major means of transportation in cities \citep{Litman2015AutonomousVI}. However, a major concern with MoD and AMoD services is that they have a tendency to become imbalanced, i.e., have a mismatch between vehicles and requests in different parts of the service area. The imbalance is due to unevenly distributed stochastic spatial and temporal travel patterns, which give rise to a poor quality of service \citep{George2012StochasticMA}. To handle this issue the service providers may match requests with vehicles centrally and proactively send vehicles to areas with predicted high imbalances. A vital part of the AMoD system is to predict the stochastic travel pattern in order to match demand \citep{Zardini_2021}. Predictions come with uncertainty in travel patterns and can have a large influence on the AMoD performance. In this paper we propose a method for efficient AMoD fleet control with probabilistic guarantees on the imbalance. The method is applied and tested for an autonomous ride-hailing service but can be applied to any MoD or AMoD system.

There are different methodologies for modeling and predicting travel demand patterns. Previous work can be divided in two categories: parametric and non-parametric travel demand prediction (in view of the structure of the demand probability distribution function). In case of parametric solutions, the underlying form of the demand is assumed to be known \citep{rasmussen}. These models span from simple linear regression to fitting of distributions to data \citep{rasmussen}. Poisson distributions are commonly used as parametric models \citep{Zardini_2021,braverman} as well as Gaussian distributions \citep{deepRL}. However, assuming such distributions are oversimplifications of demand patterns since the travel demand often follows an unknown spatio-temporal probability distribution. The complexity of the spatio-temporal travel demand patterns have lead to an increased focus on non-parametric approaches. These models do not make strong assumptions on the form of the travel demand (probability distribution), which make them more flexible to learn arbitrary patterns. Different non-parametric approaches have been used for demand prediction, such as Long Short-Term Memory neural (LSTM) networks  \citep{datadriven,SAA}. A drawback of neural network methods is that they require large data sets to be reliable. Their strong dependence on hyperparameters and initial conditions may hinder efficient fitting \citep{Pereira}. Studies on using LSTM for predicting mobility movements have been shown to be efficient with up to $80 \%$ accuracy in predicting mobility patterns \citep{lstmmob}. On the other hand, the uncertainty in the prediction should also be predicted and accounted for in the optimization, calling for explicit uncertainty parameterization. One way of modeling the uncertainty is to assume that the travel demand belongs to an uncertainty set. The uncertainty set can be constructed from data using hypothesis testing \citep{setco}. To make the uncertainty set representative enough, large data sets are required and the correlation between different time intervals is neglected. Hence, one plausible solution to explicit uncertainty estimation for forecasting spatio-temporal data with uncertainties for both small and large data sets is Gaussian Process Regression (GPR) \citep{rasmussen}. For small datasets GPR is often superior to other prediction methods and it provides a confidence on the prediction, which is beneficial for robustness, \citep{rasmussen}. However, to the best of our knowledge the efficiency of GPR is yet to be reported in combination with optimization of AMoD systems.

There has been extensive research on different modeling and control algorithms for MoD and AMoD systems. Earlier works focused on reactive control methods, from the Hungarian method \citep{Kuhn1955Hungarian} to control methods based on queueing-based models \citep{+1method,queue2}. More recently, the use of future demand together with model predictive control (MPC) have been proven to be highly effective \citep{mpcamod,datadriven,setco,SAA,9492910}. Zhang et al proposed to use MPC to solve the dispatching and rebalancing problem, \citep{mpcamod}. However, they did not consider any demand forecasting method and, in addition, the computational complexity increased with the number of vehicles. These issues were addressed in \cite{datadriven}, although the uncertainty in the demand prediction was neglected. In \cite{SAA}, a nominal prediction method is used via sample average approximation. A robust, minmax uncertainty handling is presented in \citep{setco}. \citep{MIVR} proposed a robust optimization model that combines matching of demand and vehicles with rebalancing but demand prediction was not considered. Another robust optimization model considered for vehicle rebalancing is distributionally robust optimization model with enhanced linear decision rule \citep{robustrepos}. As indicated above, there is no unique way of introducing travel demand into AMoD algorithms. Methods that take the uncertainty of the demand into account have been proven to be efficient but complex. Therefore there is a need for more transparent, scalable, computationally efficient, and accurate methods. 

A promising approach is to incorporate GPR and MPC and couple them stochastically via the uncertainty bound provided by GPR. One appealing solution is to solve the stochastic and uncertain MPC problem under probabilistic constraints, i.e. chance constraint \citep{chanceconstraint}. This methodology has proven successful for control of autonomous racing and autonomous underwater vehicles (\citep{mpcgaussianprocess}). Furthermore, chance constraint optimization (CCO) \citep{chanceconstraint} have been used in many resource allocation problems \citep{robustmpc_cco,ccodrinkingwater,VARGA202081,VARGA20181}, which are similar to the control of AMoD systems. The benefit of CCO is that, via probabilistic constraining, we can adjust the solution implicitly. This is beneficial since the two objectives of controlling a AMoD fleet, service and cost, are contradicting. Generally, the better the service, the higher the cost and vice versa. The relaxation of the CCO is typically very complex unless the probability distribution is assumed to be known. The combination of GPR and chance constrained MPC has the potential to provide a powerful methodological environment.


The main contribution of this paper is to combine data driven demand prediction with model based predictive AMoD resulting in a chance constraint optimization. This is done by first, formulating a Chance Constrained MPC (CCMPC), which is a probabilistic approach to solving stochastic optimization problems. Second, we propose a GPR for predicting travel demand time-series. The prediction given by the GPR contains both a mean prediction and an uncertainty bound. Hence, GPR naturally fits into the CCMPC framework. By means of separability and by knowing the form of the estimated PDF, probabilistic constraints can be reformulated into a deterministic optimization. To the best of our knowledge no other study has focused on the combination of GPR and CCMPC in the AMoD setting. Previous studies in this area have either focused on only the demand prediction part or on AMoD control methods, which assume simplified demand modeling or demand modeling that require large data sets. Third, the proposed optimization is benchmarked in the high fidelity transport simulator AMoDeus \citep{amodeus}. This is important in order to get an accurate measure of different metrics, such as pick-up time and vehicle mileages. Many studies consider less accurate in-house transport simulators based on simplified road and traffic models \citep{datadriven,SAA,setco}.


The outline of this article is the following. First, we present the model of the AMoD systems in Section \ref{section:modelandcontrol}. In Section \ref{section:chanceconstraint} we first formulate the MPC and then the Chance Constrained MPC, which is later relaxed to a mixed integer linear program (MILP) using GPR and the separable model. The transport simulation methodology and results are presented and discussed in Section \ref{section:results}. Finally, Section \ref{section:conclusion} concludes the paper with a short discussion and future work.











%% file: sections/model.tex
\section{AMoD Modeling} \label{section:modelandcontrol}

In this section, we first describe the stochastic discrete-time linear model of the AMoD system. This description is similar to the model presented in \citep{datadriven} and \citep{SAA}. We assume that the travel demand follows some unknown spatio-temporal probability distribution. The travel demand is predicted using GPR, which is used to relax a chance constraint model of the problem. For a detailed description of all notions introduced throughout the paper we refer to glossary in the appendix. 

\subsection{Model}
The model of the AMoD system describes the movement of vehicles and customers. It preserves vehicle and customer conservation and models the mismatch between vehicles and customers in different areas of the city. 

\begin{figure}
    \centering
    \includegraphics[width=0.4\textwidth]{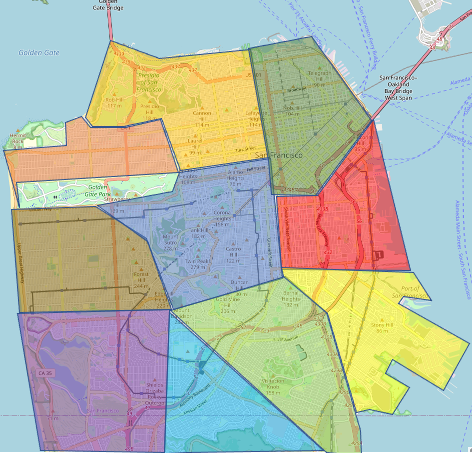}
    \caption{A visual representation of how the city of San Francisco could be partitioned into different stations \citep{OpenStreetMap}. Each colored area represents one station. }
    \label{fig:stations}
\end{figure}

The bounded operation area is a two-dimensional map denoted $\Theta \subset \mathbb{R}^2 $. We assume that the map is discretized into $N$ regions, which will be referred to as stations, (see \cref{fig:stations}), and that these partitions are given. The partitioned city is modelled as a graph network, where the nodes represent the stations and the links represent the distance between the stations. The graph is assumed to be complete, i.e. it is possible to travel in between all stations. The travel times and distances between stations are given and do not get influenced by traffic, i.e. we consider exogenous traffic. The AMoD model operates in discrete time with sample time $\Delta t$. At each time step, new customers will arrive at the stations, waiting for vehicles for pick up. The origin O and the destination D of the trip have to be within the operation area, O, D $\subseteq \Theta $. We denote the current time step as $t_0$. 

\subsubsection{States and decision variables}
There are several non-negative integer states and decision variables in the system. The first state of the system is the number of customers that want to travel from station $i$ to station $j$ at time $t$ and is denoted by $\lambda_{ij}(t)$. This is a stochastic variable and each $\lambda_{ij}(t)$ is assumed to have an unknown time-varying probability distribution, $\mathbb{P}_{ij}(t)$. The initial state of the customer demand, $\lambda_{ij}(t_0)$, is the number of outstanding customer that wants to go from station $i$ to station $j$. The second state is the average travel time in between stations and is denoted $\kappa_{ij}(t)$. The third and final state is the initial position of idle vehicles in each station and is denoted, $\phi_i(t)$. Vehicles that are traveling are assumed to be idle when they reach their destination. 

There are three decision variables in this model. The main decision variable is the movement of the vehicles when they are empty, this will be referred to as rebalancing. The number of vehicles to rebalance from station $i$ to station $j$ at time $t$ is denoted $x^{\text{r}}_{ij}(t)$. The second decision variable is the number of vehicles that serve travel demand traveling from station $i$ to station $j$ and is denoted $x^{\text{c}}_{ij}(t)$. The decision variable $s_{ij}(t)$ describes the imbalance in station $i$ for travel demand with destination $j$. The imbalance is the difference between customers and vehicles in each station. 

\subsubsection{Vehicle conservation}
Vehicle conservation means that vehicles cannot disappear nor appear in the model for a specific time period, denoted $\mathcal{T}=[1, ...,T]$ where $T$ is the number of time intervals. This is enforced through a vehicle conservation constraint, which states that the difference between vehicles entering and departing the station must be equal to the initial number of vehicles in the station for that time interval,

\begin{equation}\label{eq:flowconservation}
\begin{split}
        \sum_{j \in N}x_{ij}^{\text{c}}(t)+x_{ij}^{\text{r}}(t)-x_{ji}^{\text{c}}(t-\kappa_{ji})-x_{ji}^{\text{r}}(t-\kappa_{ji}) = \phi_i(t), \\  \qquad \forall i \in N, t \in \mathcal{T}.   
\end{split}
\end{equation}
\subsubsection{Imbalance}
The imbalance is the difference between number of travel request and vehicles in each station.
Ideally the imbalance is zero at all time, i.e. there is a perfect match between the number of travel demand and vehicles,
\begin{equation}\label{eq:imbalance}
    \lambda_{ij}(t)-x_{ij}^c(t) = 0, \\ \forall i,j \in N, t \in \mathcal{T}.
\end{equation}
However, if there are more customers than vehicles, constraint \cref{eq:imbalance} is violated. Therefore this constraint needs to be relaxed to ensure feasibility, which is done by introducing the slack variable $s_{ij}(t)$, 
\begin{equation}\label{eq:imbalancesoft}
    s_{ij}(t) = \lambda_{ij}(t)-x_{ij}^{\text{c}}(t)  \quad \forall i,j \in N, t \in \mathcal{T}.
\end{equation}
If $s_{ij}(t)>0$, i.e. there are more request then available vehicles, the remaining request should be served at a later time step. Hence, we carry on  $s_{ij}(t)$ to the next time step if $t>t_0$,
\begin{subequations}\label{eq:imbalancenew}
\begin{align}
    &\begin{aligned}
    &  s_{ij}(t+1) = s_{ij}(t)+\lambda_{ij}(t+1)-x_{ij}^{\text{c}}(t+1)  \\ &\qquad \forall i,j \in N, t \in [t_0+1,T+t_0],
    \end{aligned}\\
    & s_{ij}(t_0) = \lambda_{ij}(t_0)-x_{ij}^{\text{c}}(t_0)  \qquad \forall i,j \in N,
\end{align}
\end{subequations}

The state $x_{ij}^{\text{c}}(t)$ cannot be larger then the number of travel request since it represent only vehicles that drives customer, i.e. the imbalance should be greater or equal to zero,
\begin{align}\label{eq:geq0}
    s_{ij}(t)\geq 0, \quad \forall i,j \in N, t \in [t_0,T+t_0].
\end{align}
The combination of constraint \cref{eq:geq0} and that $s_{ij}(t)$ is a integer decision variable, gives that 
\begin{align*}
    s_{ij}(t)\in \mathbb{N}, \quad \forall i,j \in N, t \in [t_0,T+t_0].
\end{align*}

%% file: sections/problemStatement.tex
\section{Model predictive control of AMoD with probabilistic guarantees} \label{section:chanceconstraint}
In this section, a model predictive controller (MPC) for AMoD is proposed based on the model described in
the previous section. The MPC is first written as a stochastic mixed integer linear program (sMILP). The optimization problem is a MILP since several variables are restricted to be integer values \citep{walukiewicz2013integer}. The sMILP is then  reformulated as a chance constraint optimization. Finally, the Chance Constrained MPC is relaxed into a deterministic MILP using assumptions on the probability distribution of the stochastic variable.


\subsection{Model Predictive control of AMoD}
 
The above mentioned approach to AMoD is a discrete time optimization problem that is solved for $T$ time steps into the future, also called time horizon. The solution of the sMILP is a sequence of optimal decision variables for the time horizon. Only the optimal decisions belonging to the first time step in the horizon is used whilst the rest is discarded. In the next time step the states in the sMILP is updated and the sMILP is solved again, i.e. receding horizon control. The following optimization problem is formulated
\begin{subequations}\label{eq:1}
\begin{align}
        &\minimize_{x_{ij}^{\text{r}}, s_{ij}} 
        \begin{aligned}[t]
            &\sum_{t=t_0}^{T+t_0}\sum_{j,i=1}^N c_{ij}^{\text{r}}(t)x_{ij}^{\text{r}}(t) + c_{\lambda}(t)s_{ij}(t) \label{eq:1objec}
        \end{aligned} \\
        &\text{subject to} \notag \\
        & s_{ij}(t_0) = \lambda_{ij}(t_0)-x_{ij}^{\text{c}}(t_0)  \label{eq:1imb2} \qquad \forall i,j \in N,\\
        &\begin{aligned}[t]\label{eq:1imb}
            & s_{ij}(t+1)= s_{ij}(t) + \lambda_{ij}(t+1)-x_{ij}^{\text{c}}(t+1) 
         \\ &\qquad \forall i,j \in N, t \in [t_0,T+t_0], 
        \end{aligned} \\
        &\begin{aligned}[t]\label{eq:1flow} 
            &\sum_{j=1}^N x_{ij}^{\text{c}}(t)+x_{ij}^{\text{r}}(t)-x_{ji}^{\text{c}}(t-\kappa_{ji})-x_{ji}^{\text{r}}(t-\kappa_{ji}) = \phi_i(t) \\ &\qquad \forall i \in N,t \in [t_0,T+t_0],
        \end{aligned} \\
        &x_{ij}^{\text{r}},s_{ij},x_{ij}^{\text{c}}(t) \in \mathbb{N} \qquad \forall i,j \in N,t \in [t_0,T+t_0].\label{eq:1con3}  
\end{align}
\end{subequations}

Optimization problem \eqref{eq:1} is a stochastic optimal control problem. The constraints in the MPC come from the model described in the previous section, \cref{section:modelandcontrol}. The first two constraints are the imbalance in the system for $t = t_0$, \cref{eq:1imb2}, and for $t > t_0$, \cref{eq:1imb}. The third constraint is the network flow conservation, \cref{eq:1flow}. It prohibits new vehicles from appearing or disappearing from the system. The last constrain enforces the decision variables to belong to the natural numbers set, which are all non-negative integers.

The objective is to offer a good service to the customers and to ensure that this is done efficiently, \cref{eq:1objec}. Hence we want to minimize the mismatch, $s_{ij}(t)$, between customers and vehicles. This mismatch can be minimized by rebalancing vehicles in-between stations. The rebalancing comes with a cost for the operator and this cost should also be minimized. There is a trade-off between the imbalance and the rebalancing cost. This trade-off can be tuned by choosing appropriate values for the imbalance cost, $c_{\lambda}(t)$, and the rebalancing cost, $c_{ij}(t)$. The imbalance cost should reflect the cost of making customers wait, which could be varying over time. The rebalancing cost is a combination of distance and travel time. To be able to find a good weighting of the costs Pareto analysis is used.


\subsection{Chance Constrained MPC}
As mentioned in the previous section the demand prediction, $\lambda_{ij}(t)$, is assumed to follow a probability density distribution, $\mathbb{P}_{ij}(t)$. Hence, we can reformulate the imbalance constraint, \cref{eq:1imb}, to have a probability distribution fulfilled with some confidence $1-\epsilon$ where $\epsilon \in [0,1]$, see \cref{eq:ccim}. One of the benefits of this formulation is that we can decide the confidence based on what risk we want to take. Therefore the following sMILP problem is proposed
\begin{subequations}\label{eq:chanceconstraint}
\begin{align}
        &\minimize_{x_{ij}^{\text{r}}, s_{ij}} 
        \begin{aligned}[t]
            &\sum_{t=t_0}^{T+t_0}\sum_{j,i=1}^N c_{ij}^{\text{r}}(t)x_{ij}^{\text{r}}(t) + c_{\lambda}(t)s_{ij}(t) \label{eq:cc1objec}
        \end{aligned} \\
        & \text{subject to} \notag \\
        & \cref{eq:1imb2,eq:1flow,eq:1con3} \\
        &\begin{aligned}
            &\mathbb{P}_{ij}\left( s_{ij}(t+1) = s_{ij}(t) + \lambda_{ij}(t+1)-x_{ij}^{\text{c}}(t+1)\leq k\right) \\ & \geq 1-\epsilon  \qquad \forall i,j \in N,t \in [t_0,T+t_0]. \label{eq:ccim}
        \end{aligned}
\end{align}
\end{subequations}

In \cref{eq:ccim} the constant $k$ is an upper bound on the imbalance $s_{ij}(t+1)$. The Chance Constraint Optimization (CCO) problem can be difficult to solve \citep{van2011chance}. There are several methods to reformulate the chance constraints into deterministic constraints. One method is to consider that the probability distribution belongs to a set of distributions, called ambiguity sets \citep{VanParys2016}. In this work we use the separable model for reformulation of the chance constraints \citep{prekopa2013stochastic}.
\subsection{Separable Model} \label{subsec:sep model}
In the imbalance constraint, \eqref{eq:chanceconstraint}, the uncertainty and the decision variables enters in an affine way. This is a special case of the chance constraint and is referred to as a separable chance constraint \citep{shapiro}. A separable chance constraint with known probability distribution can be reformulated as a deterministic constraint. We can rewrite the separable chance constraint to a deterministic constraint by using the cumulative distribution function (CDF),
\begin{align}\label{eq:cdf}
    F_{\lambda_{ij}(t)}(z) \coloneqq \mathbb{P}_{ij}(\lambda_{ij}(t)\leq z).
\end{align}
With the use of the CDF, \cref{eq:cdf}, the chance constraint, \cref{eq:ccim}, can be written as,
\begin{align*}
 F_{\lambda_{ij}(t+1)}\left( k+x_{ij}^{\text{c}}(t+1)-s_{ij}(t)\right) \geq 1-\epsilon.
\end{align*}
Then by taking the inverse CDF we get the following constraint,

\begin{align}\label{eq:reforchance}
    k + x_{ij}^{\text{c}}(t+1)-s_{ij}(t) \geq F^{-1}_{\lambda_{ij}(t+1)}\left(1-\epsilon\right).
\end{align}
$F^{-1}_{\lambda_{ij}(t+1)}(1-\epsilon)$ is also called the quantile function. \cref{eq:reforchance} is deterministic if the CDF if known.  In this paper the travel demand, $\lambda_{ij}(t)$, is predicted using Gaussian process regression (GPR). The GPR gives a mean prediction, $\mu$, and a confidence bound on the prediction, $\sigma$, where the confidence is assumed to follow a Gaussian distribution. We can therefore use the cumulative distribution function for a Gaussian distribution, which is defined as
\begin{align} \label{eq:cdfnormal}
    F(1-\epsilon;\mu,\sigma) = \frac{1}{\sigma \sqrt{2\pi}}\int_{-\infty}^{1-\epsilon} e^{-\frac{(z-\mu)^2}{2\sigma^2}} dz.
\end{align}

Given a mean, $\mu$, and a standard deviation, $\sigma$, the cumulative distribution function is explicit, hence equation \eqref{eq:reforchance} is also explicit. The chance constraint formulation in \eqref{eq:chanceconstraint} can therefore be reformulated into the deterministic constraint \eqref{eq:reforchance}. 
\subsection{Gaussian Processes Regression (GPR) for Time-Series Modelling}


A Gaussian Process is non-parametric and probabilistic model that may be used to give predictions. GPRs are effective tools to predict time series with uncertainty bounds \citep{roberts2013gaussian}. 

The GPR can be explained from the functions perspectives, called the function-space view \citep{rasmussen}. Consider a black box system with input $\mathbf{t}$ and output $\lambda=f(t)$, where $f(t)$ is an unknown function. Assume that we have historic input- and output-data from this system, called the training data set $\mathcal{D}=\{(\mathbf{t}_i,\lambda_i)|i=1,...,n\}$. There are infinitely many functions that can be fitted on the dataset. In GPRs a probabilistic method is used to find the best function fit. This is done by assigning a multivariate probability distribution to the entire function-space. By using a probability distribution of the function space it is possible to include confidence of the prediction. 

Based on prior knowledge and a training data set the aim of GPR is to find the underlying multivariate distribution. Prior knowledge can be incorporated into the fitting process; for example periodicity or smoothness properties of $f(t)$. In GPR the underlying multivariate distribution is assumed to be a multivariate normal distribution. Hence the estimated output follows a normal distribution, $\lambda_1,...,\lambda_n \sim \mathcal{N}(\mu(\mathbf{t})_{i,..,n},\mathbf{\Sigma})$, where $\mathbf{\Sigma_{i,j}}= $ Cov$(\lambda_i,\lambda_j)=k(t_i,t_j)$ is the covariance function, also called kernel,  and $\mu(\mathbf{t})$ is the mean function. Thus, the Gaussian process is completely defined by its mean and covariance functions according to
\begin{equation}
    f(t) \sim \mathcal{GP}(\mu(\mathbf{t}),k(\mathbf{t},\mathbf{t}')).
\end{equation}
An important aspect of kernels is that they are only dependent on the inputs. The covariance function can be any function that generates a positive semi-definite covariance matrix \citep{bookGP}. When selecting different kernels, prior knowledge of the data is used. If we assume a smooth function the radial basis function kernel (RBF) can be used
\begin{equation}
    k_{\text{RBF}}(\mathbf{t},\mathbf{t}') = \exp{\bigg(-\frac{\|\mathbf{t}-\mathbf{t}'\|^2}{2l^2}\bigg)},
\end{equation}
where $l$ is the lengthscale hyperparameter. If the data is periodic a periodic kernel is proposed
\begin{equation}
    k_{\text{Periodic}}(\mathbf{t},\mathbf{t}') = \exp{\bigg(-2\frac{\sin^2{(\frac{\pi }{p}}(\mathbf{t}-\mathbf{t}'))}{l^2}\bigg)},
\end{equation}
where $p$ is the period and $l$ the lengthscale hyperparameter. The sum and multiplication of two kernels is also a kernel \citep{rasmussen}.  

When the kernels have been selected the hyperparameters are trained on the dataset by maximizing the log-marginal likelihood \citep{rasmussen}. The log marginal likelihood is given by
\begin{equation}
\label{eq:logmarg}
    \log p(\mathbf{y}|X,\theta) = -\frac{1}{2}\mathbf{y}^\top \mathbf{\Sigma}^{-1}\mathbf{y}-\frac{1}{2}\log|\mathbf{\Sigma}|-\frac{n}{2}\log(2\pi),
\end{equation}
where $\mathbf{\Sigma}$ is the covariance matrix,
\begin{equation}
\mathbf{\Sigma_{n,n}} = 
\begin{pmatrix}
k_{1,1} & k_{1,2} & \cdots & k_{1,n} \\
k_{2,1} & k_{2,2} & \cdots & k_{2,n} \\
\vdots  & \vdots  & \ddots & \vdots  \\
k_{n,1} & k_{n,2} & \cdots & k_{n,n} 
\end{pmatrix}.
\end{equation}
A gradient method is used to find the hyperparameters that maximizes the log marginal likelihood, i.e. the partial derivatives of \cref{eq:logmarg} with respect to the hyperparameters are computed:
\begin{equation}
    \frac{\partial}{\partial \theta_i} \log p(\mathbf{y}|X,\theta) = -\frac{1}{2}\mathbf{y}^\top \mathbf{\Sigma}^{-1}\frac{\partial \mathbf{\Sigma}}{\partial \theta_i}\mathbf{\Sigma}^{-1}\mathbf{y}-\frac{1}{2}\text{tr}(\mathbf{\Sigma}^{-1}\frac{\partial \mathbf{\Sigma}}{\partial \theta_i})
\end{equation}
The computational complexity of training the GPR is mainly due to the need of finding the matrix inversion of $\mathbf{\Sigma}$ and requires the computational complexity ($\mathcal{O}(n^3)$) \citep{rasmussen}.  
From the tuned kernels and mean function, future prediction can be made using conditional probability on the posterior distribution. Given a new input $t^*$, the predictive distribution of the corresponding output $\lambda^*$ is a Gaussian distribution with mean and variance 
\begin{flalign}
\centering
\hat{\mu}(t^*) & = \mathbf{k_*}^\top\mathbf{\Sigma}^{-1}\mathbf{y}, & \label{eq:mean} \\
\hat{\sigma}^2(t^*) & = k(t_*,t_*)-\mathbf{k_*}^\top\mathbf{\Sigma}^{-1}\mathbf{k_*}, & \label{eq:std}
\end{flalign}

where $\mathbf{\Sigma}$ is the covariance matrix for the training data, $\mathbf{k_*}$ is the vector of covariances between $t^*$ and the n training points.  
\begin{figure}[!htp]
    \centering
    \includegraphics{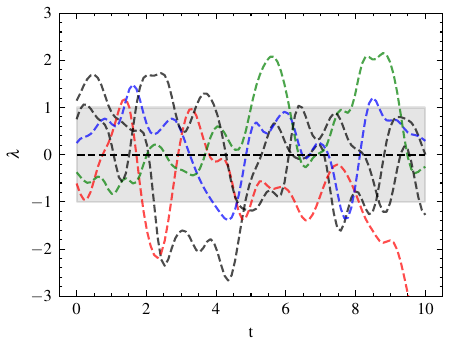}
    \caption{Samples from prior distribution of the locally periodic kernel.}
    \label{fig:locallyperiodic}
\end{figure}
In this work we will use a locally periodic kernel which is the multiplication of the RBF and Periodic kernels. Periodic kernels assumes perfect correlation between data points that are $N*p$ distances apart, i.e. $\mathbf{t}-\mathbf{t}'=N*p$, where $N$ is an integer. This strict periodicity assumption is not valid for most stochastic functions. While travel demand data have some periodicities they are not strict, e.g., the exact time and extend of people commuter patterns varies from day to day. Locally periodic kernels allow the shape of the periodic parts to vary over time and are therefore better suitable for travel demand prediction. Arbitrary function samples from the prior of the locally periodic kernel can be seen in \cref{fig:locallyperiodic}. It can be seen that there are local periodicity in each sample but that the periodicity can change over time.




\subsection{Chance Constraint MPC (CCMPC) with GPR}
The chance constraint in optimization \cref{eq:chanceconstraint} can be reformulated to an deterministic constraint using the separable model in \cref{subsec:sep model}. The mean and standard deviation in equation \cref{eq:cdfnormal} is estimated using GPR, \cref{eq:mean,eq:std}. Hence the chance constraint represent the confidence in the prediction of the travel demand. The estimated mean and standard deviation are denoted $\hat{\mu}$ and $\hat{\sigma}$, \cref{eq:mean,eq:std}.  The final optimization problem can be written as,
\begin{subequations}\label{eq:chanceconstraintgpr}
\begin{align}
        &\minimize_{x_{ij}^{\text{r}}, s_{ij}} 
        \begin{aligned}[t]
            &\sum_{t=t_0}^{T+t_0}\sum_{j,i=1}^N c_{ij}^{\text{r}}(t)x_{ij}^{\text{r}}(t) + c_{\lambda}(t)s_{ij}(t) \label{eq:cc1gprobjec}
        \end{aligned} \\
        & \text{subject to} \notag \\
        & \cref{eq:1imb2,eq:1flow,eq:1con3} \\
&\begin{aligned}
    &s_{ij}(t+1)+k+x_{ij}^{\text{c}}(t+1)-s_{ij}(t) \\ & \qquad \geq F^{-1}_{\lambda_{ij}(t+1)}\left(1-\epsilon;\hat{\mu},\hat{\sigma}\right),  \\ & \qquad \forall i,j \in N,t \in [t_0,T+t_0], \label{eq:ccgprim}
\end{aligned} \\
\end{align}
\end{subequations}

The chance constraint in \cref{eq:ccim} is reformulated to \cref{eq:ccgprim} using the separable model \cref{eq:reforchance}. The optimization problem, \cref{eq:chanceconstraintgpr}, is now an deterministic MILP. An important property with \cref{eq:chanceconstraintgpr} is that the MILP is totally unimodular and hence the corresponding linear program (LP) solution will always be integral. Therefore, the optimization problem \cref{eq:chanceconstraintgpr} can be solved efficiently as a LP with the simplex method. The proof that \cref{eq:chanceconstraintgpr} is totally unimodular can be found in Appendix \ref{app:TU}.
\subsection{Minimal fleet size}

The chance constraint \cref{eq:ccim} guarantees that the imbalance in each station $i$ is below a threshold $k$ with probability $1-\epsilon$. However, this guarantee is only valid if we have enough vehicles in the station to drive the predicted demand, $x_{ij}^{\text{c}}(t)$. The decision variable $x_{ij}^{\text{c}}(t)$ is constrained by \cref{eq:1flow}. Hence, we need to rebalance vehicles between station and ensure that the total fleet size is large enough. The minimal fleet size can be found by solving the following optimization problem,

\begin{subequations}\label{eq:minimalfleet}
\begin{align}
        &\minimize_{x_{ij}^{\text{r}}, \phi_i(0)} 
        \begin{aligned}[t]
            &\sum_{t=t_0}^{T+t_0}\sum_{j,i=1}^N c_{ij}^{\text{r}}(t)x_{ij}^{\text{r}}(t) + c_{\phi}\phi_i(0) \label{eq:minfleet1}
        \end{aligned} \\
        & \text{subject to} \notag \\
        &\begin{aligned}
            &k + x_{ij}^{\text{c}}(t+1) \geq F^{-1}_{\lambda_{ij}(t+1)}\left(1-\epsilon;\hat{\mu},\hat{\sigma}\right) \\ & \qquad \forall i,j \in N,t \in [t_0,T+t_0], \label{eq:minfleet2}
        \end{aligned} \\
        &\begin{aligned}[t]\label{eq:minfleet3} 
            &\sum_{j=1}^N x_{ij}^{\text{c}}(0)+x_{ij}^{\text{r}}(0)-x_{ji}^{\text{c}}(-\kappa_{ji})-x_{ji}^{\text{r}}(-\kappa_{ji}) = \phi_i(0) \\ &\qquad \forall i \in N,
        \end{aligned}\\
        &\begin{aligned}[t]\label{eq:minfleet4} 
            &\sum_{j=1}^N x_{ij}^{\text{c}}(t)+x_{ij}^{\text{r}}(t)-x_{ji}^{\text{c}}(t-\kappa_{ji})-x_{ji}^{\text{r}}(t-\kappa_{ji}) = 0 \\ &\qquad \forall i \in N, t \in [1,T],
        \end{aligned} \\
        &\cref{eq:1con3}.
\end{align}
\end{subequations}
The optimization problem minimize the rebalancing, $x_{ij}^{\text{r}}$, and the initial number of vehicles in each station, $\phi_i(0)$. We optimize for an full-day horizon and the imbalance in each station is set to zero, $s_{ij}(t)$.

\subsection{Algorithm}
\begin{figure*}[!htp]
    \centering
    \includegraphics[width=\textwidth]{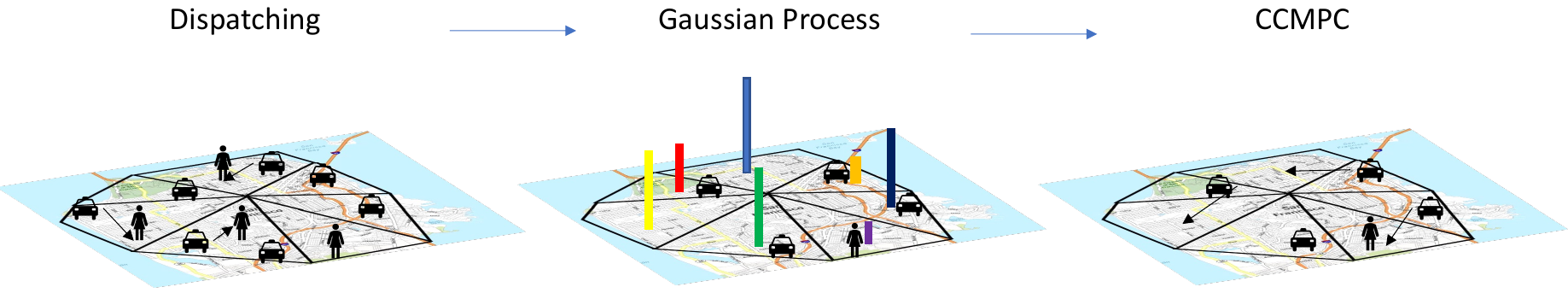}
    \caption{Visual representation of the dispatching, travel demand prediction and MPC. The map is split into $N$ stations and there is an initial number of requests and vehicles. First, request are matched with vehicles in the same station. Then the travel demand is predicted, here represented by the colored bars. Based on the travel demand prediction the AMoD optimization problem is solved (\cref{alg:one}) and the vehicles are rebalanced accordingly.}
    \label{fig:algorithm}
\end{figure*}

The proposed algorithm for 1 time step is presented in \cref{alg:one}. The control actions are updated every $\Delta t_{MPC}$ minutes and the GPR is updated every $\Delta t_{GP}$ minutes. For dispatching of request and vehicles in each station we use the Hungarian algorithm \citep{Kuhn1955Hungarian}. It is important to note that the Hungarian can only match request and vehicles in the same station. The proposed \cref{alg:one} make sure that there are enough vehicles in each station to serve the request. Before the dispatching and rebalancing is started the city/rural area is discretized into $N$ stations using $k$-means clustering on historical request location. A visual representation of the dispatching, prediction and rebalancing can be seen in \cref{fig:algorithm}.


\RestyleAlgo{ruled}

\SetKwComment{Comment}{/* }{ */}

\begin{algorithm}[!htp]
\caption{AMoD dispatching and rebalancing}\label{alg:one}
\textbf{Input} Time t, System state ${\phi_i(t)}$, Historical demand $\lambda_{ij}^{\text{hist}}$, Currently waiting customer $\lambda_{ij}^{\text{wait}}$, probability $1-\epsilon$. \\
 
\textbf{Output:} Control action $x^{\text{r}}$. \\
\eIf{$t\mod\Delta t_{GP}$}{
    Train GPR \\
  Obtain $\{\hat{\mu}(t),\hat{\sigma}(t)\}_{t \in [t_0,T+t_0]}$
    }{Use previous GPR \\
    Obtain $\{\hat{\mu}(t),\hat{\sigma}(t)\}_{t \in [t_0,T+t_0]}$
}
\If{$t\mod \Delta t_{\text{MPC}}$}{
$x^{\text{r}} \gets \text{solve}$ \quad \cref{eq:chanceconstraintgpr} 
}
\Return{ $x^{\text{r}}$}
\end{algorithm}

%% file: sections/simulationResults.tex
\section{Case Study} \label{section:results}
In this section, the proposed Chance Constrained MPC algorithm outlined in \cref{alg:one} is tested in realistic AMoD scenarios using a high-fidelity transport simulator. 

\subsection{Simulation Environment}
The high-fidelity transport simulator  AMoDeus \citep{amodeus} was used for contrasting and benchmarking of the AMoD algorithm (see \cref{alg:one}). AMoDeus is an open-source agent-based transport simulator based on Multi-Agent Transport Simulator (MATSim) \citep{matsim}. It was intentionally developed to simulate AMoD systems and to test new algorithms for fleet control. With AMoDeus and MATSim large scale transport simulations for one full day can be performed. The transport network is constructed using a queue based approach. Several realistic mobility scenarios for different cities are implemented in AMoDeus for benchmark testing. We chose to simulate the San Francisco scenario for this work. The San Francisco scenario is based on a taxi data set from 2008 \citep{epfl-mobility-20090224}. The data contains mobility traces from 500 taxi vehicles in San Francisco and contains 464 045 customer trips, which were collected between May 17, 2008 and June 10, 2008. In this study we have chosen to simulate Thursday May 29th, which corresponds to a total of 11 453 requests. The transport simulations were performed on a MacBook Pro with a 2.3 GHz Quad-Core Intel core i7 processor and 16 GB of RAM. We used IBM CPLEX to solve the ILP problems \citep{cplex2009v12}.

\subsection{Travel Demand Prediction}

The Gaussian Process regression is trained using the python library GPyTorch \citep{GPyTorch}. GPyTorch is an efficient implementation of GPRs that uses PyTorch for GPU acceleration. This enable us to train the GPRs with high speed, which enables real-time usages. 

\begin{figure}
    \centering
    \includegraphics{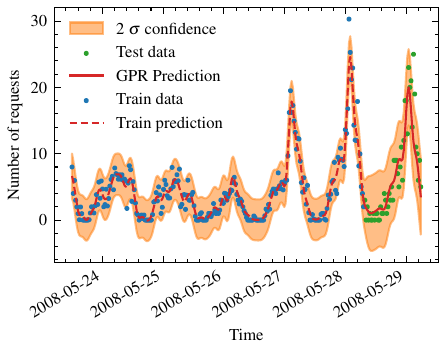}
    \caption{Gaussian Process regression of requests going from one station to another. The blue data points indicates the training data while the green data points corresponds to the request we want to predict. The red line is the prediction of incoming requests and the orange area is the corresponding $95\%$ confidence bound.}
    \label{fig:gpr}
\end{figure}

The training data is the flow of requests in between all stations per time interval. Hence, a GPR is trained for each specific flow. For this case we have discredited the city into 10 stations and hence we need to train 100 GPRs \footnote{Optimal partitioning or dynamic repartitioning is out of the scope of this paper. However, we experimented with the granularity and found 10 districts adequate. Adequate in the sense of having enough data to train traveler demand with GP.}. The training data consist of the previous 5 days from the day we want to predict, in this case we use training data from 2008-05-24 to 2008-05-28. The trained mean and $95\%$ confidence for the flow between station 0 and 3 can be seen in \cref{fig:gpr}. The mean prediction and the confidence describes the data well. We can see in the peak on day 2008-05-29 that the mean prediction is low but the confidence includes the peak test data points. The explained variance score and mean squared error (MSE) for all trained GPRs can be seen in \cref{tab:gprmetric}. The mean explained variance score is low and with a relative high standard deviation. This indicates that the selected kernel is not good for predicting all flows, which is expected because of randomness in travel patterns. However, since we account for prediction uncertainty in \cref{alg:one}, we explicitly can handle it. Also, the mean of the MSE metric is quite low which indicates that our average behavior is adequate. 

The computational complexity of GPR is cubic in the number of data points ($\mathcal{O}(n^3)$) \citep{rasmussen}. This makes GPRs inadequate for large dataset. However, GPyTorch reduces the computational complexity to be squared in the number of data points ($\mathcal{O}(n^2)$). This alongside the use of GPU hardware results in an acceptable computation burden for our case studies. For this work the average computational training time per prediction for GPR was 4.82 seconds, see \cref{tab:compMPC}. If the city is split into 10 stations it means that one hundred different flows have to be predicted. Hence the total computational time will be around 482 seconds. 
\begin{table}
\centering
\caption{\label{tab:gprmetric}Metric scores for GPRs.}
\begin{tabular}{lll}
Metric                   & Mean  & Standard Deviation  \\ \hline
Explained Variance Score & 0.362 & 0.225               \\
Mean Squared Error       & 2.054  & 5.151             
\end{tabular}
\end{table}
\subsection{Minimal fleet size for different confidence level}\label{sec:minfleet}
By solving the optimization problem \cref{eq:minimalfleet} we get the minimal required fleet size to keep the imbalance in each station below some threshold, $k$, with probability $1-\epsilon$.
\begin{figure}
    \centering
    \includegraphics{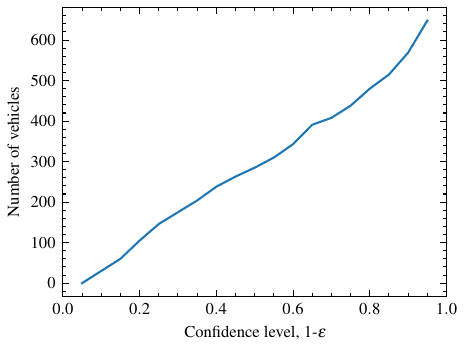}
    \caption{The minimal fleet size, optimal solution from \cref{eq:minimalfleet}, for different confidence $1-\epsilon$.}
    \label{fig:minfleet}
\end{figure}

Intuitively the minimal required fleet size should increase with higher probability, $1-\epsilon$, which is also the case see \cref{fig:minfleet}. The results in \cref{fig:minfleet} is from solving \cref{eq:minimalfleet} with threshold $k=0$. When $1-\epsilon$ is increasing the bound on the travel demand, $\lambda_{ij}(t)$, increases and hence a larger fleet size is required. 
\subsection{Confidence bound in CCMPC}

\begin{figure}
    \centering
    \includegraphics{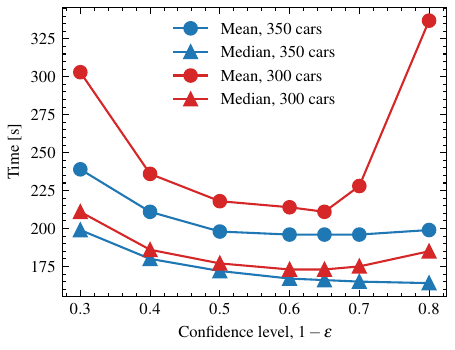}
    \caption{Mean and median wait time for different confidence bounds for fixed fleet size of 300 and 350 vehicles using CCMPC \cref{alg:one}.}
    \label{fig:epsmeanandmedian}
\end{figure}

\begin{figure}
    \centering
    \includegraphics{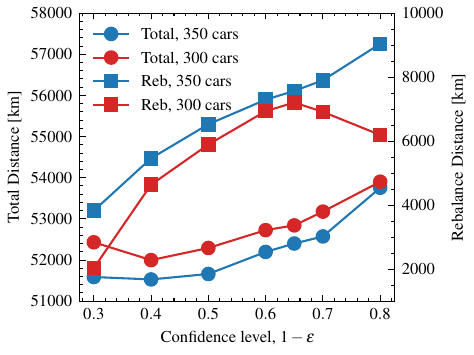}
    \caption{Total and rebalancing distance for different confidence bounds for fixed fleet size of 300 and 350 vehicles.}
    \label{fig:epsdist}
\end{figure}

In the CCMPC (\cref{alg:one}) there are four variables that can be tuned according to the specific scenario; the objective cost weights ($c_{ij}^{\text{r}}(t),c_{\lambda}$), the horizon length $T$, the confidence bound $1-\epsilon$ and the fleet size. We have chosen to have a horizon of 3 hours and the cost weights are set according to operating costs, where the rebalancing cost is relative to the distance and the imbalance cost is relative to customer wait time. With a fixed horizon and cost weights we study how the confidence bound in the chance constraint optimization problem (\cref{eq:chanceconstraint}) affects the performance for fleet sizes of 300 and 350 vehicles. Two performance metrics are evaluated; the pick-up time, see \cref{fig:epsmeanandmedian}, which indicates the service level provided, and the distances driven, see \cref{fig:epsdist}, which represents operating cost. 

By studying the mean and median pickup times as a function of confidence level for 300 vehicles it is evident that there is an optimal confidence level at around 0.65 where both of these measures are minimized (see \cref{fig:epsmeanandmedian}). The maximum mean wait time is reached for a confidence level of 0.8 and the second-highest for a confidence level of 0.3. When the confidence level is increased, the number of predicted requests between the different station's increases. Hence, a higher confidence level require a larger fleet size, which we concluded in \cref{sec:minfleet}. For a fleet size of 300 vehicles there is not enough control input for confidence level of 0.8 and the rebalancing hence decreases, see \cref{fig:epsdist}. When the vehicle fleet is increased to 350 vehicles there is possibility for more rebalancing and the rebalancing increases for confidence level 0.8. However, the performance in terms of median and mean pick-up time is similar or worse then for confidence level 0.65. The extra rebalancing is not improving the service level since it rebalance more then it have too.
For low confidence levels the number of requests are underestimated, hence the rebalancing decreases for lower confidence levels (see \cref{fig:epsdist}). Even though the rebalancing distance is the lowest for confidence level 0.3, the total distance is lowest for confidence level 0.4 which has more than the doubled rebalancing distance compared to confidence level 0.3 (see \cref{fig:epsdist}). Beyond 0.4, the distance grows due to stricter and stricter chance constraints forcing vehicles to drive towards customers. A confidence level of 0.65 have the maximal rebalancing distance of 7206 km and the total distance is only 844 km more than the minimum total distance for fleet size of 300 vehicles. This indicates that the rebalancing at confidence level 0.65 decreases the mean and median wait time at a low cost. Hence, this confidence level is considered to be optimal for this scenario. For this confidence level the median wait time is reduced by $4\%$ compared to using only the mean prediction of the GPR.





\subsection{Comparative evaluation of AMoDs}
The performance of the CCMPC (\cref{alg:one}) is benchmarked against three different control algorithms:
\begin{itemize}
    \item\emph{MPC-Oracle} - A non-causal controller where the future travel demand, $\lambda_{ij}(t)$, is known for all $t$ in \cref{eq:1}. I.e. the performance of this controller is an upper limit for the performance of the proposed algorithm.
    
    \item\emph{MPC-FixedDemand} - This is a causal controller, see \cref{eq:1}, with a fixed future travel demand, all future travel demands are set equal to the last known travel demand, i.e., $\lambda_{ij}(t)=\lambda_{ij}(t_0-1) \, \forall \, t \in [t_0,T+t_0] $.
    
    \item \emph{Global Bipartite Matching Dispatcher (GBM)} - This controller solves the bipartite problem to match available vehicles with customer request using the Hungarian algorithm \citep{Kuhn1955Hungarian}. The cost of matching a request with a vehicle is the distance between them. The controller does simply react to the current demand and does not perform any rebalancing.
\end{itemize}

\begin{figure}[ht]
    \centering
    \includegraphics{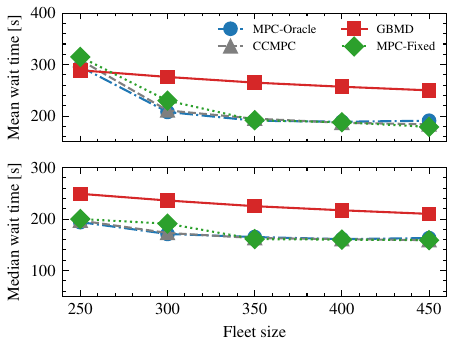}
    \caption{Mean and median wait times as function of fleet size for different control algorithms.}
    \label{fig:fleetwait}
\end{figure}

\begin{figure}[ht]
    \centering
    \includegraphics{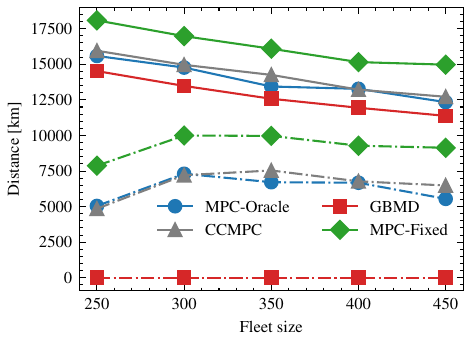}
    \caption{Driving distance as function of fleet size for different control algorithms. Solid lines corresponds to empty distance, i.e. distance driven without customers, and dash-dotted lines corresponds to rebalance distance.}
    \label{fig:fleetdist}
\end{figure}

A comparison of the performance, pickup time and distance driven, for different control algorithms as a function of fleet size can be seen in \cref{fig:fleetwait,fig:fleetdist}. It is apparent that the best performing algorithm in terms of wait time is the MPC-Oracle, which is expected. However, for a fleet size of 300 vehicles the best performing causal algorithm is the CCMPC with only a few seconds more mean and median pickup time (see \cref{fig:fleetwait}). For a fleet size of 300 vehicles the mean pickup time for CCMPC is $24\%$ lower then the GBM (see \cref{tab:waittime}), which is the worst performing algorithm in terms of pickup time. This is also expected since GBM is a reactive algorithm. The MPC-fixed is performing well in terms of pickup time and have a 19 seconds longer mean pickup time compared to CCMPC (see \cref{tab:waittime}). However, the MPC-Fixed have a rebalancing distance that is 2764 km more than CCMPC. Since the fixed demand prediction is not accurate a lot of vehicles will unnecessarily be rebalanced. When the fleet size is above 300 vehicles the performance of MPC-Oracle, CCMPC and MPC-fixed in terms of pickup time is similar because a large fleet size compensates for a bad controller as the MPC-Fixed. A lot of vehicles can be rebalanced without affecting the pickup-time since there is an oversupply of vehicles in the system. Therefore, the mean and median pickup-time is similar but the total distance driven is still more for the MPC-Fixed. From an operator's perspective, it is desired to keep the fleet size as low as possible because of cost savings.  Therefore, a fleet size of 300 vehicles seems to be optimal in terms of cost and performance (see \cref{fig:fleetwait,fig:fleetdist}). 

\subsection{Computational Complexity}
MILP is an NP-hard problem, meaning that the time required for optimizing does not scale well and cannot be solved in polynomial time. However, 
since the MILP is totally unimodular the solution of the relaxed LP is always integral. Therefore, we can solve the optimization problem, \cref{eq:chanceconstraintgpr}, as a LP. The mean computational running time is 0.093 seconds, see \cref{tab:compMPC}.  An important aspect of the CCMPC (\cref{alg:one}) is that the computational complexity is not dependent on the number of vehicles nor the number of requests. The computational complexity is only dependent on the number of stations, which makes the CCMPC suitable for control of large scale vehicle fleets. 

\begin{table}[ht]
\caption{\label{tab:waittime}Pickup times, total-,rebalance- and pickup-distances for fixed fleet size of 300 vehicles and each control algorithm.}
\resizebox{0.45\textwidth}{!}{\begin{tabular}{lllll}
Metric      & MPC Oracle & CCO MPC & GBM & MPC Fixed \\ \hline
Mean pickup time {[}s{]}   & 205        & 211     & 276       & 230          \\
Median pickup time {[}s{]} & 171        & 173     & 236       & 191    \\
Total distance {[}km{]}       & 52645   & 52842 & 51361  &  54858         \\
Rebalance distance {[}km{]} & 7307    & 7206  & 0         &  9970         \\
Pickup distance {[}km{]}      & 7464    & 7757  & 13482  & 6983     
\end{tabular}}
\end{table}

\begin{table}[]
\caption{\label{tab:compMPC}Computational running time for single execution of Chance Constrained MPC (CCMPC) for routing and Gaussian Process Regression (GPR) for travel demand prediction.}
\resizebox{0.45\textwidth}{!}{\begin{tabular}{llllll}
Method & Samples & Mean {[}s{]} & Median {[}s{]} & STD {[}s{]} & Max {[}s{]} \\ \hline
CCMPC & 143     & 0.093        & 0.078         & 0.047
                     & 0.281      \\
GPR & 886 & 4.82        &      4.82    &                 0.112     &       6.87
\end{tabular}}
\end{table}

%% file: sections/conclusion.tex
\newpage
\section{Conclusion and Future Work} \label{section:conclusion}

In this paper, we have proposed a predictive chance constraint rebalancing approach for autonomous mobility-on-demand (AMoD) services, which is applied to the use case of ride-hailing. We first introduce a commonly used model for this service where the service area is discretized into smaller areas, called stations. The model consist of constraints for the imbalance and vehicle conservation. Based on the model a model predictive controller (MPC) is formulated with the multi-objective to minimize the rebalance distance for vehicles and the imbalance in each station. The travel demand is predicted using Gaussian Process regression (GPR). GPR, in contrast to other proposed prediction methods, is superior for small data sets and provides a confidence bound on the prediction. We account for uncertainties in the travel demand prediction by formulating a chance constraint MPC (CCMPC). The CCMPC is relaxed using the GPR prediction and the use of the separable model. 
The proposed algorithm was benchmarked using the high fidelity transport simulator AMoDeus and real taxi data from San Francisco \citep{amodeus}. Our results show the importance of incorporating the confidence bound of the demand prediction. By tuning the confidence bound the median wait time is reduced by $4\%$ compared to using only the mean prediction of the GPR. We showed that the CCMPC is performing close to optimal performance and that and is significantly better than a reactive controller.
The performance and computational efficiency of the proposed method implies that it would be useful for real-time control. There are many important directions to consider for future work including embedding traffic and limited range into the model as well as more case studies for different cities.


%% file: sections/acknowledgment.tex
\section*{Acknowledgment}
This work was co-funded by Vinnova, Sweden through the project: Simulation, analysis and modeling of future efficient traffic systems, by the FFI program Efficient and Connected Transport Systems. Balázs Kulcsár is acknowledging the partial support of the Area of Advanced Transport.